\documentclass[10pt,journal,compsoc,onecolumn]{IEEEtran}
\usepackage{cite}
\usepackage[pdftex]{graphicx}
\graphicspath{{../pdf/}{../jpeg/}}
\DeclareGraphicsExtensions{.pdf,.jpeg,.png}
\usepackage{array}
\usepackage{amsmath}
\usepackage{amssymb}
\usepackage{mdwmath}
\usepackage{mdwtab}
\usepackage{eqparbox}
\usepackage{url}
\usepackage{hyperref}
\usepackage[switch]{lineno}
\usepackage{xcolor}

\usepackage{tcolorbox}
\tcbset{standard jigsaw, opacityback=0.5}

\begin{document}
% \linenumbers
\title{Automatic Organ and Pan-cancer Segmentation in Abdomen CT: the FLARE 2023 Challenge}

\author{Jun Ma, Yao Zhang, Song Gu, Cheng Ge, Ershuai Wang, Qin Zhou, Ziyan Huang, Pengju Lyu, Jian He, and Bo Wang % <-this % stops a space
\IEEEcompsocitemizethanks{\IEEEcompsocthanksitem Jun Ma is with the Department of Laboratory Medicine and Pathobiology, University of Toronto; Peter Munk Cardiac Center, UHN AI Hub, University Health Network; Vector Institute, Toronto, Canada\\
Yao Zhang is with AI Lab, Lenovo Research, Beijing, China\\
Song Gu is with the Department of Image Reconstruction, Nanjing Anke Medical Technology Co., Ltd., Nanjing, China\\
Cheng Ge is with Ocean University of China, Qingdao, China\\
Ershuai Wang is with Department of Research and Development, ShenZhen Yorktal DMIT Co. LTD., Shenzhen, China \\
Qin Zhou is with Institute of Medical Robotics, School of Biomedical Engineering, Shanghai Jiao Tong University; Department of Computer Science and Engineering, East China University of Science and Technology, Shanghai, China\\
Ziyan Huang is with Shanghai Jiao Tong University; Shanghai AI Laboratory, Shanghai, China\\
Pengju Lyu is with City University of Macau, Macau, China; Hanglok-Tech Co., Ltd., Hengqin, China\\
Jian He is with the Department of Nuclear Medicine, Nanjing Drum Tower Hospital,  Nanjing, China \\
Bo Wang is with the  Peter Munk Cardiac Center, University Health Network; Department of Laboratory Medicine and Pathobiology and Department of Computer Science, University of Toronto; Vector Institute; UHN AI Hub, University Health Network, Toronto, Canada\protect\\
\IEEEcompsocthanksitem Corresponding author: Bo Wang.
E-mail: bowang@vectorinstitute.ai}% <-this % stops an unwanted space
% \thanks{Manuscript received Jan. 10, 2023}
}

\IEEEtitleabstractindextext{%
\begin{abstract}
Organ and cancer segmentation in abdomen Computed Tomography (CT) scans is the prerequisite for precise cancer diagnosis and treatment. Most existing benchmarks and algorithms are tailored to specific cancer types, limiting their ability to provide comprehensive cancer analysis. 
This work presents the first international competition on abdominal organ and pan-cancer segmentation by providing a large-scale and diverse dataset, including 4650 CT scans with various cancer types from over 40 medical centers. The winning team established a new state-of-the-art with a deep learning-based cascaded framework, achieving average Dice Similarity Coefficient (DSC) scores of 92.3\% for organs and 64.9\% for lesions on the hidden multi-national testing set. The dataset and code of top teams are publicly available, offering a benchmark platform to drive further innovations \url{https://codalab.lisn.upsaclay.fr/competitions/12239}.
\end{abstract}

}

\maketitle

\IEEEdisplaynontitleabstractindextext
\IEEEpeerreviewmaketitle

\section*{Introduction}\label{intro}
Abdomen organs are quite common cancer sites, such as colorectal cancer and pancreatic cancer, which are the second and third most common cause of cancer death~\cite{CancerStatis22}. Computed Tomography (CT) scanning yields important prognostic information for cancer patients and is a widely used imaging technology for cancer diagnosis and treatment monitoring~\cite{PancreasCT-NMed}. In both clinical trials and daily clinical practice, radiologists and clinicians measure the tumor and organ on CT scans based on manual measurements (e.g., Response Evaluation Criteria In Solid Tumors (RECIST) criteria)~\cite{RECIST2009}. However, this manual assessment is inherently subjective with considerable inter- and intra-expert variability and cannot measure the 3D tumor morphology.

Deep learning-based methods have shown great potential for automatic tumor segmentation and quantification. Many challenges have been established to benchmark algorithm performance by providing standard datasets and fair evaluation platforms, such as the brain tumor segmentation (BraTS)~\cite{bakas2018brats}, liver and liver tumor segmentation~\cite{LiTS-MIA}, kidney and kidney tumor segmentation~\cite{KiTS2021MIA}, and pancreas and colon lesion segmentation~\cite{MSD-Summary}. These challenges have greatly advanced algorithm development~\cite{LiverLesionSeg-MICCAI23,nnunet21-NM}, but they only focus on one type of lesion (e.g., liver cancer, kidney cancer, or pancreas cancer), which cannot provide holistic lesion analysis. 
Pan-cancer segmentation in abdomen CT plays an important role in clinical practice because lesions can spread from one organ to another organ. For example, pancreas cancer and colorectal cancer could transfer to the liver, leading to liver metastases. There is a great need for general algorithms that can segment all kinds of lesions from CT scans.

% P3. current progress on 
Recently, universal lesion segmentation algorithms for abdomen CT have received increasingly attention~\cite{unilesionSeg-MIA, uni-lesionseg-miccai, unilesionSeg-ICCV, unilesion-det-segICCV23}. 
However, these algorithms are developed and evaluated under various datasets, leading to difficulties in fairly comparing them. The main barrier is the lack of a general benchmark platform and publicly available dataset. 
In this work, we addressed the limitation by providing the largest abdominal pan-cancer dataset and organized the first international competition to prompt the development of universal abdominal organ and lesion segmentation algorithms.  In particular, we curated a diverse abdominal pan-cancer dataset with 4650 CT scans, covering various abdomen cancers from 50 medical centers, which is the most comprehensive abdomen pan-cancer dataset to date. The competition attracted 292 participants from all over the world. 
The winning algorithm surpassed existing state-of-the-art models and achieved average DSC scores of $92.31\% \pm 3.3$ and $64.9\% \pm 27.4$ for organ and lesion, respectively. 
The inference pipeline consumed less than 4 GB GPU memory with an average runtime of $8.58 \pm 1.92$ seconds, which can be deployed on consumer desktops.

\begin{figure*}[!htbp]%
\centering
\includegraphics[scale=0.36]{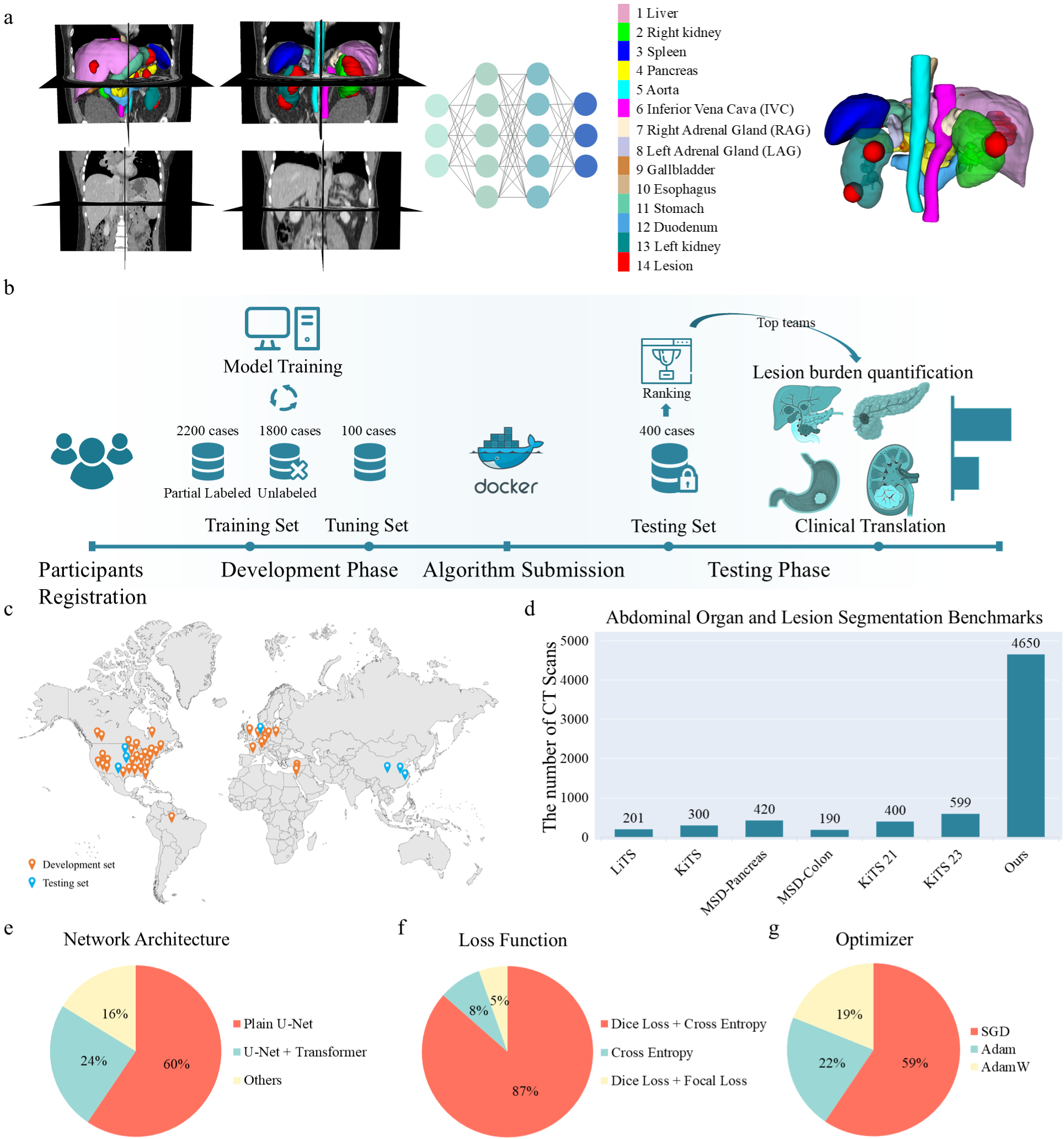}
\caption{\textbf{Overview of the challenge design.} \textbf{a}, The challenge task is to automatically segment 13 abdominal organs and all kinds of lesions in abdomen CT. \textbf{b}, The complete challenge pipeline. \textbf{c}, Data source distributions of the challenge dataset. \textbf{d}, Comparison to existing abdominal organ and lesion segmentation benchmarks in terms of the number of CT scans.  Distribution of key designs among 37 algorithms: \textbf{e}, network architecture, \textbf{f}, loss function, and \textbf{g}, optimizer.
}\label{fig:overview}
\end{figure*}

\section*{Results}\label{sec2}
\subsection*{Challenge design}

This challenge aimed to prompt the methodology development of fully automatic abdominal organ and lesion segmentation algorithms in CT scans. 
Different from the previous FLARE challenges designed for pure organ segmentation~\cite{FLARE21-MIA, FLARE22-LDH}, the FLARE 2023 challenge introduced three improvements (\ref{fig:overview}a). 
First, the challenge task was expanded to joint organ and lesion segmentation, including 13 abdominal organs and one general lesion class. Notably, the lesion class focused on abdominal pan-cancer segmentation rather than a single type of cancer. 
Second, we increased the dataset size from 2050 to 4000 CT scans, the largest abdomen CT benchmark, for model development. 
Third, we formulated the challenge as a partially supervised learning task instead of fully supervised learning because only parts of organs or lesions are labeled in clinical practice, and annotating all regions of interest is expensive. 

During the past decade, abdominal organ and lesion segmentation and partial-label learning methods have received increasing attention, but these approaches are developed with different datasets and evaluation metrics, which has led to difficulties in fair comparisons in various studies~\cite{ImperfectData-MIA,partial-organ-lesion-PAMI}.
This challenge is set up in a timely way to bridge the gap.
The challenge design followed the Biomedical Image Analysis ChallengeS (BIAS)~\cite{BIAS}, which has been pre-registered at the 26th International Conference on Medical Image Computing and Computer Assisted Intervention (MICCAI 2023)~\cite{MA-FLARE23-Design}\footnote{https://conferences.miccai.org/2023/en/challenges.asp}. We also provided the BIAS~\cite{BIAS} and CLAIM (Checklist for Artificial Intelligence in Medical Imaging)~\cite{CLAIM} checklists in the supplementary.

The challenge consisted of two phases (Fig.~\ref{fig:overview}b). During the development phase, participants received 2200 partially labeled cases and 1800 unlabeled cases CT scans for model training. We also provided 100 fully labeled cases for model tuning and set up a public leaderboard where participants can submit segmentation masks and compare the performance with others. Each team can submit up to three results on the leaderboard per day. 
During the testing phase, we held a hidden testing set with 400 fully labeled cases and one of their reference standard was publicly available. Each team was required to submit their algorithm via the docker container and we manually ran the docker container on the same workstation to generate the testing set segmentation results. 
To avoid overfitting the testing set, each team can only submit one algorithm docker container.

\subsection*{Dataset characteristic}
CT scans are very diverse in different medical centers because of various imaging protocols, manufacturers, and diseases. 
We curated the challenge dataset by aggregating CT scans from multiple medical centers (Fig.~\ref{fig:overview}c, Supplementary Table 1-5) and all the datasets were allowed to be used and shared for research purposes.  The training and tuning sets were mainly from North America and Europe while the testing set contained unseen centers from Asia, aiming to evaluate the generalization ability of the submitted algorithms. 
Moreover, the dataset covered all CT phases, such as the plain phase, artery phase, portal phase, and delay phase, as well as common CT manufacturers, including GE Healthcare, Philips, Siemens, and Toshiba. 
Fig.~\ref{fig:overview}d shows the statistics of the number of CT scans in existing abdominal tumor segmentation challenges. Our dataset is 23 times larger than the widely used LiTS challenge dataset and substantially exceeds the existing largest abdominal tumor segmentation dataset (KiTS23).

\begin{figure*}[!htbp]%
\centering
\includegraphics[scale=0.35]{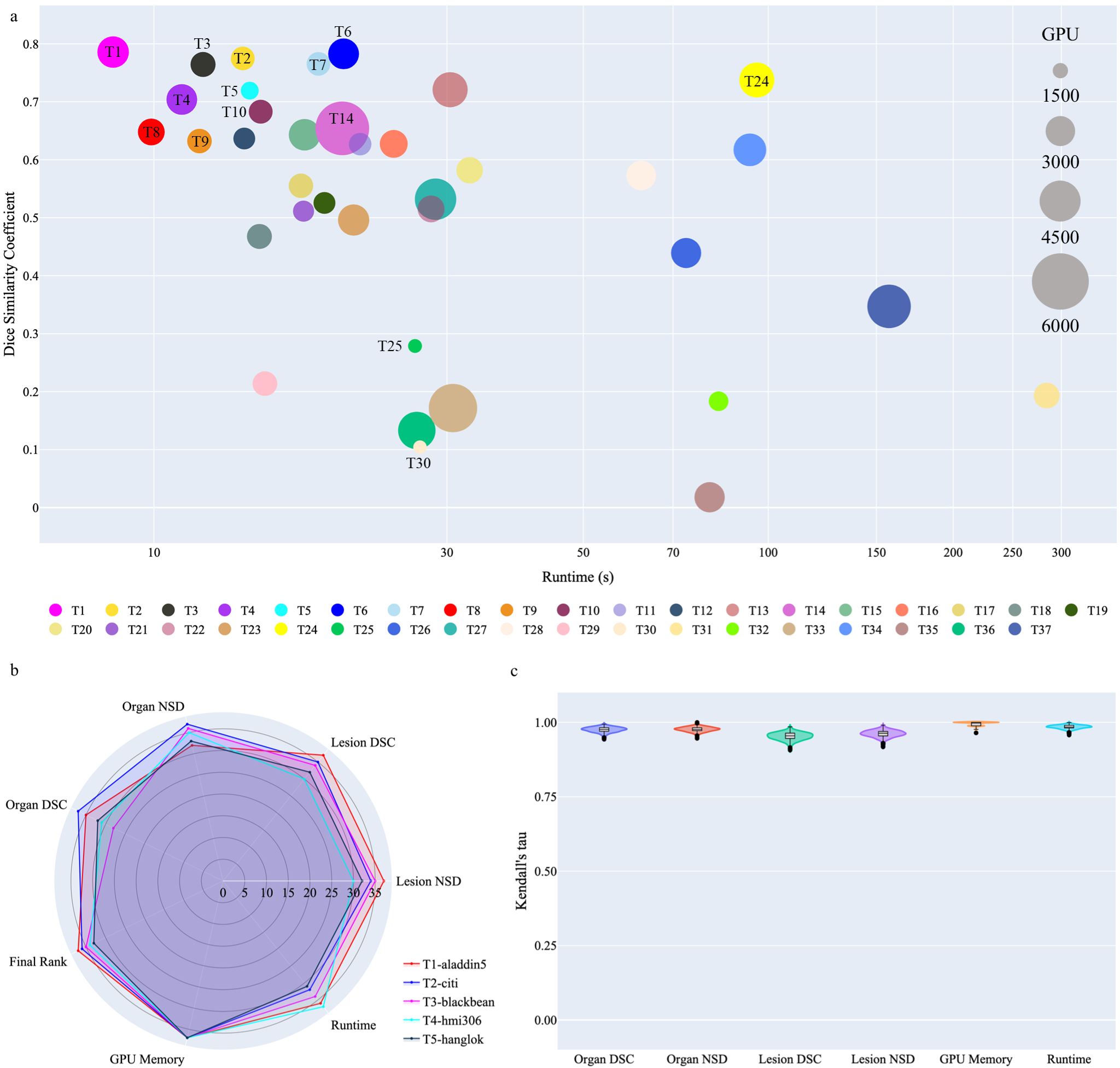}
\caption{\textbf{Segmentation accuracy and efficiency performance analysis on the testing set (N=400).} \textbf{a}, Segmentation accuracy  (\textit{y} axis) and efficiency (x axis) results of all the 37 algorithms. Each circle denotes one team and the circle size is proportional to the GPU memory consumption. The top algorithms are on the top left with a better tradeoff. \textbf{b}, Performance comparison of top five teams across all the metrics. Each color indicates one team and the value denotes the number of algorithms it surpassed on the corresponding metric.  \textbf{c}, Ranking stability analysis results for all metrics by bootstrap approach (number of samples $b=1000$). The violin plot visualizes the distribution of Kendall’s $\tau$ values with a central box plot embedded to show the interquartile range, median, and outliers. The overall consistency of high Kendall's $\tau$ values across the metrics underscores a stable performance evaluation of the algorithms across different dimensions.
}\label{fig2:results}
\end{figure*}

\subsection*{Overview of evaluated algorithms}
47 teams from 292 participants joined the challenge and we received 37 successful algorithm docker container submissions during the testing phase, where four submissions failed and the remaining six teams did not submit. 
We analyzed three key components of the employed deep learning model among the 37 teams, including network architectures, loss functions, and optimizers. All teams used 3D networks and 60\% of the teams used U-Net~\cite{nnunet21-NM} as the main architecture. The combination of Dice loss and cross-entropy loss was the most popular loss function, used by 87\% of the teams. Stochastic gradient descent was usually used for optimizing the U-Net while Transformer-based networks usually used Adam~\cite{Adam} and its variant (AdamW~\cite{AdamW}).

\begin{figure*}[!htbp]%
\centering
\includegraphics[scale=0.25]{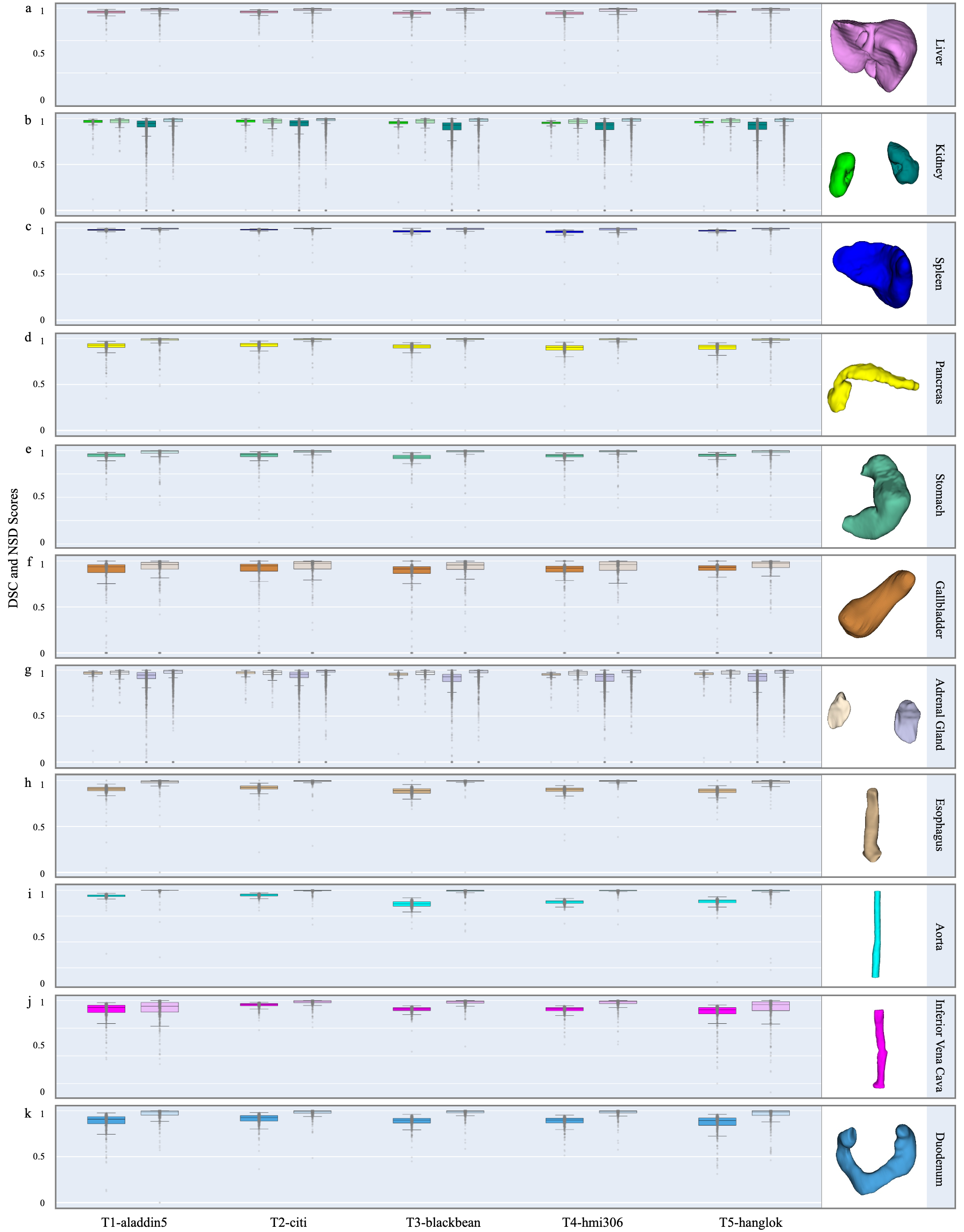}
\caption{\textbf{Organ-wise segmentation results of the top five teams on the testing set (N=400).} The box plots display the average DSC (deep color) and NSD (light color) scores for each organ across all testing cases, with the median value represented by the black horizontal line within the box, the lower and upper quartiles delineating the borders of the box, and the vertical black lines indicating the 1.5 interquartile range.  
}\label{fig3:organ}
\end{figure*}

\textbf{Best-performing algorithm.} 
Team aladdin5 (T1~\cite{FLARE23-1st}) designed an efficient cascaded framework that localized the region of interest first followed by fine-grained segmentation for organs and lesions. The organ pseudo labels of the unlabeled organs, generated by the best-accuracy-algorithm~\cite{aladdin5-Flare22} in FLARE 2022~\cite{FLARE22-LDH} were used to enlarge the annotations in the training set. 
In particular, a lightweight nnU-Net~\cite{nnunet21-NM} was first trained on the combination of labeled cases and unlabeled cases with pseudo labels for binary segmentation of the region of abdominal organs. After that, two individual models were trained with the cropped images with different image spacing for organ and tumor segmentation, respectively. The segmentation models were further fine-tuned with selected patches where the organ or tumor was centered. To improve inference efficiency, the prediction interpolation was implemented on GPU and multithreading was employed to pre-process images and load model checkpoints simultaneously.

\textbf{Second-best-performing algorithm.} 
Team citi (T2~\cite{FLARE23-2nd}) presented a partially supervised framework with nnU-Net~\cite{nnunet21-NM} to leverage the partially labeled data. During partially supervised training, the images were first grouped by their annotated classes and then a batch of training images were selected from one group with a probability of its proportion. In this way, the images in a training batch should have identically labeled classes. When calculating the loss between the prediction and the ground truth, the output channels of unlabeled classes were merged to one channel by max pooling and corresponded to the background channel in the reference standard. The pseudo labels were selected by uncertainty estimation and further cleaned by filtering out small isolated regions. Due to the lack of tumor annotations, a CutMix augmentation strategy was exploited to copy tumor regions from the cases with labeled tumors to those without labeled tumors. To preserve the context of each tumor region, the neighboring regions were cropped together with the tumor object. 

\textbf{Third-best-performing algorithm.} 
Team blackbean (T3~\cite{FLARE23-3rd}) developed a self-training framework that exploits a large model for pseudo label generation and a small model for efficient segmentation. Specifically, a large STU-Net-L~\cite{huang2023stu} was trained on 250 cases with complete organ annotations and generated pseudo labels for 1497 cases with tumor annotations. A fully annotated dataset with 1497 cases can be obtained by merging the tumor and pseudo organ labels. Then, another STU-Net-L was trained on the 1497 cases to generate both organ and tumor pseudo labels for the remaining cases. Finally, a small STU-Net-B was trained on the whole dataset with complete manual or pseudo labels. Inference pipeline was optimized by applying large target spacing, avoiding the cropping process, and using GPU-based interpolation for fast image resampling.

\textbf{Fourth-best-performing algorithm.} 
Team hmi306 (T4~\cite{FLARE23-4th}) proposed a two-stage pipeline that first located the abdominal organs and then segmented organs and tumors respectively. In the first stage, PHTrans~\cite{liu2022phtrans}, a hybrid network consisting of convolutional network and Swin Transformer~\cite{liu2021swin}, was trained for binary organ segmentation in low-resolution images. The second stage integrated self-training and mean teacher to leverage the unlabeled and partially labeled data for organ and tumor segmentation respectively. For organ segmentation, a PHTrans model was trained on fully annotated abdominal organ data followed by generating organ pseudo labels for unlabeled data. Then, another PHTrans model was trained on both labeled and pseudo labels. For tumor segmentation, following the mean teacher approach~\cite{tarvainen2017mean}, a student model is supervised by the prediction of a teacher model and the reference standard. The teacher model is updated by exponential moving average of trained student models. Both teacher and student models employ a lightweight ResU-Net and a whole-volume-based input strategy for highly efficient segmentation.

\textbf{Fifth-best-performing algorithm.} 
Team hanglok (T5~\cite{FLARE23-5th}) introduced a cascaded framework with a Transformer-based architecture and self-training. A lightweight binary segmentation network with partially convolutional encoder~\cite{chen2023run} and SegFormer~\cite{xie2021segformer} decoder was first trained to localize the ROI in low-resolution images. Then, a Transformer-based segmentation network with MetaFormer~\cite{yu2022metaformer} encoder and UNETR~\cite{hatamizadeh2022unetr} decoder was used to segment the 13 organs and tumor from the ROI in high-resolution images. To enhance spatial context and reduce computation cost, the conventional self-attention~\cite{vaswani2017attention} was replaced by depth-wise convolution with different kernel sizes and a group-wise convolution for multi-scale aggregation. To leverage the unlabeled data, the winning algorithm in FLARE22~\cite{blackbean-Flare22-Design} was used to initialize the pseudo labels of unlabeled data. Then, self-training was adopted to refine the pseudo labels and optimize the segmentation model.

\subsection*{Segmentation accuracy and efficiency analysis on the testing set}
We show the testing set segmentation accuracy and efficiency performance of the 37 teams in Fig.~\ref{fig2:results}a (Supplementary Table 6). The majority of teams are concentrated on the top left of the plot, indicating that most participants aimed to develop accurate and efficient algorithms. However, we also noticed that some teams only pursue the unilateral metric. For example, T25 and T30 consumed around 1500MB GPU memory, but the segmentation accuracy is inferior to the others. T24 achieved very competitive DSC score but the inference speed is slow, costing 99 seconds for each case.  
In contrast, the winning team (T1) stands out with an average DSC of 92.3\% and 64.9\% for organ and lesion segmentation, respectively, and an inference speed of 8.6s by consuming 3561.6 MB of GPU RAM.

Next, we compared the performance of top five algorithms across seven metrics (organ DSC and NSD, lesion DSC and NSD, runtime, GPU memory consumption, and final rank) in terms of the number of algorithms it outperformed, aiming to provide a comprehensive comparison of the algorithms' strengths and weaknesses across the multiple criteria (Fig.~\ref{fig2:results}b). The radar plot reveals that the top five algorithms outperformed most others across these metrics, as evidenced by the significant overlap in the plotted areas. However, closer inspection highlights subtle variations in performance. In particular, T1-aladdin5 demonstrated the highest accuracy in lesion segmentation, while T2-citi excelled in organ segmentation. They all obtained the perfect GPU consumption metric, but T4-hmi406 achieved the fastest inference speed. Overall, T1-aladdin5 secured the best final rank with a better balance between segmentation accuracy and computational efficiency.

We also analyzed the ranking stability of all the employed metrics regarding testing case sampling variability. Specifically, the bootstrap approach was used by generating 1000 bootstrap samples where each sample contained 400 randomly selected testing cases with replacement from the testing set. Then, we compute Kendall's $\tau$ values for all the metrics and Fig.~\ref{fig2:results}c shows the corresponding distributions with violin and box plots. The Kendall's $\tau$ values for all metrics are clustered around 1.0, implying that the rankings of the algorithms are highly consistent and stable. 

\subsection*{Organ-wise performance analysis}
We further present a detailed analysis of the performance of top five teams in segmenting a diverse set of 13 organs (Fig.~\ref{fig3:organ}, Supplementary Table 7). Specifically, we categorized these organs into three groups based on their size, morphology, and segmentation challenges: large solid organs, small organs, and tubular organs. This classification allows us to highlight the specific difficulties and successes associated with each category, providing insights into how the algorithms perform differently across varying anatomical structures.

The first group consists of large solid organs, including the liver, kidneys, spleen, pancreas, and stomach. These organs generally have well-defined structures, making them more straightforward targets for segmentation algorithms. The liver, kidneys, and spleens demonstrated high and consistent performance in all five teams, with average DSC and NSD scores above 95\% in all five teams (Fig.~\ref{fig3:organ}a-c). 
Their box plots also show tight clusters, suggesting that these organs are easier for segmentation due to the relatively large size and distinct boundaries. The pancreas, despite being part of this group, exhibited greater variability with average DSC scores ranging from 88.9\% to 91.3\%, reflecting the anatomical complexity and variability in shape and size across patients, which presents additional challenges for accurate segmentation (Fig.~\ref{fig3:organ}d). The stomach also performs well, with mean DSC scores between 93.4\% (T4) and 94.9\% (T1), indicating overall good segmentation accuracy across the teams (Fig.~\ref{fig3:organ}e).

The second group focuses on small organs, specifically the gallbladder and adrenal glands (Fig.~\ref{fig3:organ}f-g). These organs are characterized by their smaller size and less distinct boundaries, making them more difficult to segment. The DSC and NSD scores exhibit greater variability than large organs, ranging from 83.5\% (T1-aladdin5) to 86.9\% (T4-hmi306). The adrenal glands show a similar trend, with mean DSC scores from 79.0\% (T3-blackbean) to 89.5\% (T2-citi).
These results suggest that the small size and less distinct boundaries of these organs pose substantial challenges for accurate and robust segmentation, leading to increased variability in performance across the different algorithms.

The final group includes tubular organs, such as the esophagus, aorta, inferior vena cava (IVC), and duodenum, which are defined by their elongated, tube-like shapes, which pose unique challenges for segmentation. The performance across this group was more diverse, particularly for the duodenum and esophagus.
Specifically, the aorta and inferior vena cava have high and consistent accuracy with mean DSC scores ranging from 90.6\% (T5-hanglok) to 98.0\% (T2-citi) (Fig.~\ref{fig3:organ}i-j). In contrast, the esophagus shows more variability, with mean DSC scores ranging from 87.8\% (T5-hanglok) to 91.6\% (T2-citi) (Fig.~\ref{fig3:organ}h). The duodenum, another challenging tubular organ, has mean DSC scores ranging from 86.1\% (T5-hanglok) to 90.8\% (T2-citi), indicating the difficulty in accurately segmenting this organ due to its elongated and irregular shape (Fig.~\ref{fig3:organ}k).

\subsection*{Lesion performance analysis}
The challenge task approached the lesion as a semantic segmentation problem, categorizing each voxel as either part of a lesion or not. This approach aligns with organ segmentation tasks, enabling uniform evaluation across different cases. Alternatively, the task can be framed as an instance segmentation problem, which not only captures category information but also distinguishes between different lesions. This allows for the identification and analysis of multiple distinct lesions within the same image.

\begin{figure*}[!htbp]%
\centering
\includegraphics[scale=0.2]{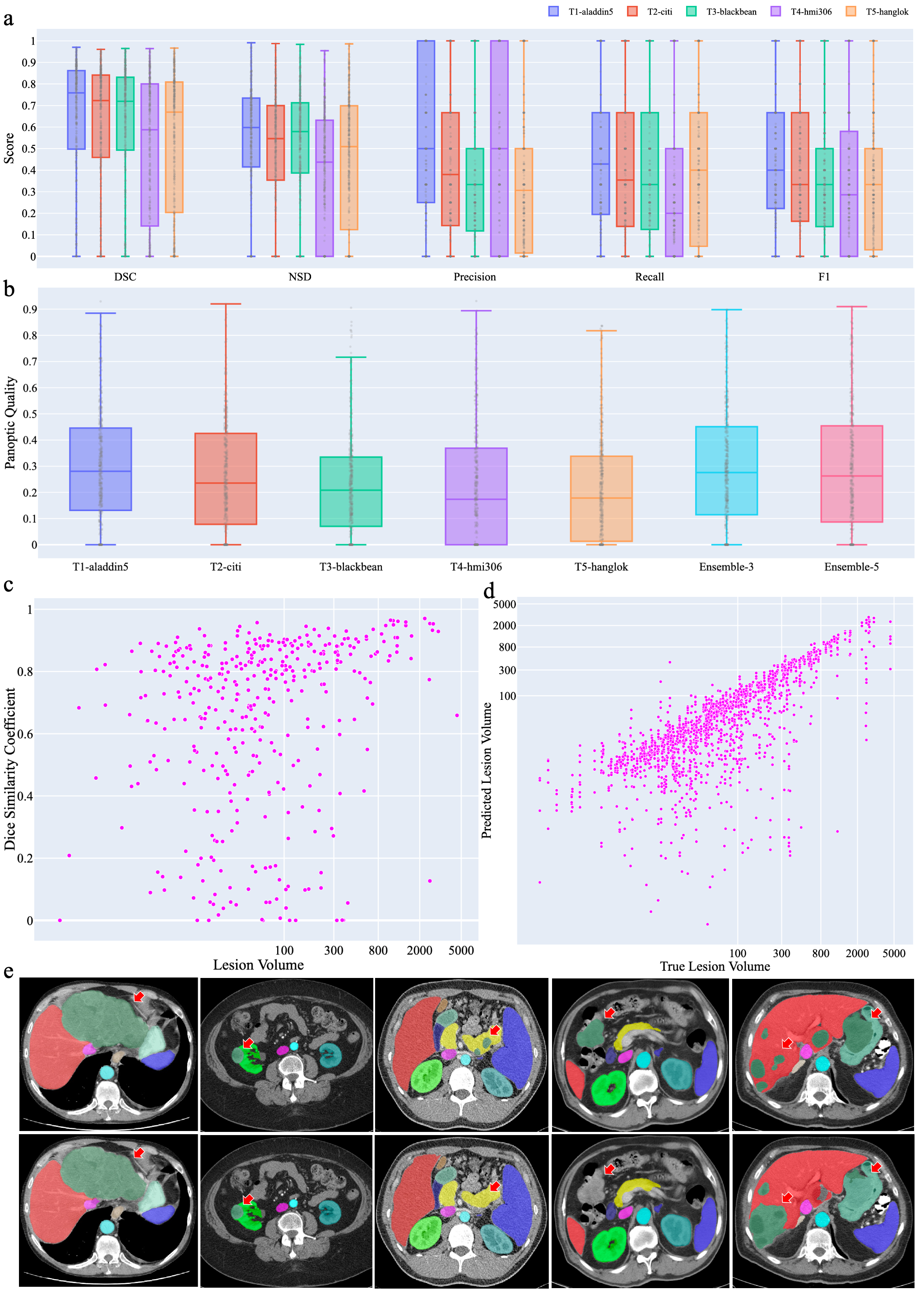}
\caption{\textbf{Lesion semantic and instance segmentation results of the top five algorithms on the testing set (N=400).} \textbf{a}, Dot-box plots for lesion semantic segmentation metrics (DSC and NSD scores) and instance segmentation metrics (Precision, Recall, and F1 scores). \textbf{b}, Lesion panoptic quality of the top five algorithms and the ensemble of top three and top five algorithms. 
\textbf{c}, Relationship between segmentation accuracy and lesion volume of the winning algorithm.
\textbf{d}, Relationship between the predicted and true lesion volume of the winning algorithm.
\textbf{e}, Visualized organ and tumor segmentation examples of the winning algorithm. 
The top row shows the reference standards and the bottom row shows the segmentation results.
}\label{fig4:lesion}
\end{figure*}

We evaluated the lesion segmentation results of the top five teams using both semantic segmentation metrics (DSC and NSD) and instance segmentation metrics (Sensitivity, Specificity, and F1 score), where we separated the disconnected lesions as individual entities with connected components analysis. The dot-box plots show the testing set performance distribution for each team across the five metrics (Fig.~\ref{fig4:lesion}a, Supplementary Table 8). All the top three teams achieved a median DSC score of over 70\% where T1-aladdin5 had the highest median DSC score of 75.9 (interquartile range (IQR):49.8-86.2\%). However, the median NSD scores dropped below 60\% for all the teams, indicating that lesion boundaries may not be accurately delineated and small lesions could be missed.

In terms of instance segmentation metrics, all the teams achieved high precision but low recall. This imbalance indicates that the algorithms are conservative in their segmentations. In particular, T1-aladdin5 and T4-hmi306 obtained the best specificity with median scores of 50.0\% (IQR: 25.0-100.0\%) and 50.0\% (IQR: 0.0-100.0\%), respectively. However, T4-hmi306 had the lowest sensitivity with a median score of 20.0\% (IQR: 0.0-50.0\%), indicating that many lesions were missed in the segmentation results.
The F1 scores follow a similar trend to the DSC scores with T1-aladdin5 achieving the highest median score of 40.0\% (IQR: 22.2-66.7\%). Overall, the F1 scores across teams show a broad range, reflecting inconsistencies in balancing lesion detection precision and recall.

We also employed the majority vote approach to generate the ensemble results of the top three and top five algorithms, respectively, and computed the panoptic quality metric to understand both segmentation and detection quality.
As shown in Fig.~\ref{fig4:lesion}b (Supplementary Table 9), the ensemble models generally show comparable or slightly improved panoptic quality compared to the individual teams, with Ensemble-3 slightly outperforming the others in terms of median score. However, all teams and ensembles exhibit a wide range of panoptic quality scores, indicating variability in performance across different cases. 

Next, we analyzed the relationship between lesion volume and segmentation accuracy (DSC) of the winning algorithm T1-aladdin5 (Fig~\ref{fig4:lesion}c).
There is a clear trend where larger lesion volumes tend to correspond with higher DSC scores, suggesting that the algorithm performs better as the lesion volume increases. For smaller lesions, particularly those with volumes below approximately 100, the DSC values are widely dispersed, ranging from near 0 to around 80\% or higher. This indicates that the segmentation algorithm struggles more with smaller lesions, leading to less consistent and generally lower DSC scores.

Tumor volume is an important image biomarker and we compared the predicted lesion volume to the true lesion volume (Fig.~\ref{fig4:lesion}d). The result reveals that while the model generally performs well in predicting lesion volumes, with a strong overall correlation between true and predicted volumes, it exhibits variability in accuracy, particularly at the extremes of the volume range. Smaller lesions tend to be underestimated, while larger lesions are occasionally overestimated, as indicated by the spread of data points around the diagonal line in the scatter plot.

Finally, we present a visualization of typical examples (Fig~\ref{fig4:lesion}e) that contains two successful cases (the 1st and 2nd columns) and three failure examples (the 3rd to 5th columns). These examples demonstrate that the algorithm is capable of accurately identifying and segmenting both large and small lesions when their appearances and boundaries are well-defined. However, the algorithm often struggles with small, heterogeneous lesions, as evidenced by its complete failure to detect the pancreas and colorectal lesions leading to poor performance.

\section*{Discussion}\label{diss}
% P1. Existing Unsolved problems and study objectives and hight advantages of this challenge
AI has revolutionized medical image segmentation tasks, but most algorithms rely on a large number of human expert annotations, which are extremely hard and expensive to collect. Moreover, the performance of existing algorithms is mainly evaluated based on accuracy-related metrics on limited cohorts while the generalization ability, running efficiency, and resource consumption are overlooked. These barriers hinder the wider adoption of AI algorithms in clinical practice. The main goal of this study was to address these critical issues. In particular, we created the largest abdomen organ and pan-cancer CT dataset with a well-defined segmentation task to benchmark algorithms. Furthermore, we organized an international competition to gather community efforts to facilitate algorithm development with partially unlabeled data. All the algorithms were validated in real-world settings by blinded evaluation of their generalization ability, efficiency, and resource consumption on intercontinental and multinational cohorts.

The final results of the FLARE 2023 challenge reveal four main findings. 
First, despite the increased difficulty posed by the inclusion of lesion segmentation, U-Net-based algorithms~\cite{nnunet21-NM}, as demonstrated by the top-performing teams, continued to achieve the highest lesion segmentation accuracy without compromising organ segmentation performance. This outcome underscores the robustness of the U-Net model, confirming its capability to effectively handle both general abdominal organ and lesion segmentation tasks.

Second, we have identified some useful strategies to enhance segmentation accuracy. For example, the cascaded segmentation framework, beginning with region of interest (ROI) extraction followed by fine-grained segmentation, allows the model to focus on the segmentation targets and extract detailed information for better accuracy. 
Additionally, the use of a weighted compound loss function, which combines Dice loss and focal loss, proved beneficial in addressing class imbalance, particularly improving the segmentation of smaller organs and tumors. Further accuracy improvements were achieved through strategic training methods, including selective patch sampling, fine-tuning with optimized learning rates and batch sizes, and model ensembling to leverage the strengths of different models. A post-processing step, which retains the largest connected component, was also recommended to boost organ segmentation accuracy.

Third, top-performing teams also implemented several strategies to boost segmentation efficiency. These included utilizing GPU-based operations for faster probability interpolation and label generation, replacing slower CPU-based preprocessing with GPU processing, and adopting multi-threading to minimize model initialization times. Together, these strategies contributed to the development of robust and efficient segmentation models, well-suited to the low-resource hardwares of clinical applications.

Fourth, the utility of unlabeled data in improving organ segmentation accuracy was demonstrated across all top teams. However, its impact on lesion segmentation was less pronounced, likely due to the inaccuracy of lesion pseudo labels, which could have introduced negative effects during iterative training. While the top algorithm surpassed existing state-of-the-art models~\cite{unilesionSeg-ICCV,flare23-zongwei,flare23-Andriy}, none of the teams or ensemble approaches achieved high segmentation quality for lesion segmentation, highlighting areas for further improvement in both segmentation accuracy and detection robustness.

This work has two primary limitations. First, the dataset contained a relatively small number of lesion annotations. Second, the challenge focused exclusively on abdominal lesions, whereas other cancer types, such as lung cancer, are also critical in clinical practice. Future work could address these limitations by expanding the dataset to add synthetic lesion images~\cite{LiverLesionsyn-cvpr23, lesionsyn-cvpr24} and other publicly available abdomen CT datasets~\cite{amos-data,AbdomenAtlas}, as well as cancer types~\cite{deeplesion}.
Extending the challenge to include multimodal data is another promising direction, given that text data has shown potential in enhancing lesion detection and segmentation accuracy.~\cite{lesion-multimodal,unilesionSeg-MedIA}.

In summary, the FLARE 2023 challenge presents the first and largest benchmark for organ and pan-cancer segmentation in abdomen CT. The winning algorithm set a new state of the art with the cascaded framework and efficient network and inference pipeline designs. The top algorithms have achieved high accuracy for most of the organs but small organs and lesions remain unsolved problems. All the data and code have been made publicly available for further algorithm developments.

\section*{Methods}\label{method}
\subsection*{Challenge schedule} 
The FLARE 2023 challenge was preregistered~\cite{MA-FLARE23-Design} and the proposal passed peer review at the 25th International Conference on Medical Image Computing and Computer-Assisted Intervention (MICCAI 2023). 
We launched the challenge on April 1st 2023 on the CodaLab platform~\cite{codalab}. During the development phase, each team can submit up to three tuning set segmentation results every day to the online platform and get the segmentation accuracy scores. 
Moreover, each team also had five chances to submit docker containers to challenge organizers and obtain segmentation efficiency scores. 
During the testing phase, participants were required to submit the final algorithm docker by August 25th 2023. We manually evaluated all the submitted dockers on the hidden testing set and announced the results on October 8th 2023 at MICCAI.

\subsection*{Data standardization and annotation protocol}
The data standardization followed the common practice in the other 3D medical image segmentation challenges~\cite{LiTS-MIA,MSD-Summary,KiTS2021MIA} and the past FLARE challenges~\cite{FLARE21-MIA,FLARE22-LDH}. We curated the CT scans from public datasets based on the license permission. Detailed information on these datasets is presented in Supplementary Table 1-4. 
All CT scans were converted to the standard NIfTI format (\url{https://nifti.nimh.nih.gov/}) and preserved the original CT HU values. 
The orientation was standardized as canonical 'RAS', which means that the first, second, and third voxel axes go from left to Right, posterior to Anterior, and inferior to superior, respectively. Organ annotation protocol remained the same as for the FLARE 2022 challenge~\cite{FLARE22-LDH}, which adhered to the radiation therapy oncology group consensus (RTOG) panel guideline~\cite{goodman2012RTOG} and Netter's anatomical atlas~\cite{netter-atlas2014}. In the training set, the lesion annotations in the source datasets were directly used. In the tuning and testing set, all visible lesions were annotated by a senior radiologist with the assistance of ITK-SNAP~\cite{ITKSNAP} and MedSAM~\cite{MedSAM}.

\subsection*{Evaluation protocol}
All algorithms were sequentially run on the same GPU desktop workstation for a fair evaluation. The workstation was a Ubuntu 20.04 desktop with one central processing unit (CPU, Intel Xeon(R) W-2133 CPU, 3.60GHz × 12 cores), one graph processing unit (GPU, NVIDIA QUADRO RTX5000, 16G), 32G of memory, and 500G of hard disk drive storage. 
We used two groups of metrics to evaluate segmentation accuracy and efficiency. 
Following the recommendations in Metrics Reloaded~\cite{MetricsReloaded}, the segmentation accuracy metrics contained Dice Similarity Coefficient (DSC) and Normalized Surface Distance (NSD), measuring the region and boundary overlap between segmentation mask and reference standards. 
The segmentation efficiency metrics included runtime and Area Under the Curve of GPU memory-time (AUC GPU), where the GPU memory consumption was recorded every 0.1s. 
In addition, we also analyzed commonly used instance segmentation metrics for lesion segmentation, including precision, recall, F1 score, and panoptic quality.

\subsection*{Ranking scheme}
The final rank was computed with both segmentation accuracy and efficiency metrics. We give runtime and GPU memory consumption a tolerance of 15s and 4GB, respectively, because they are acceptable in clinical practice. The employed metrics cannot be directly merged because of the dimension difference. Thus, we used rank-then-aggregation to obtain the final rank. Specifically, the ranking scheme had three steps:
\begin{itemize}
    \item Step 1. Compute the six metrics for each case in the testing set (N=400), including two organ-wise metrics: average DSC and NSD scores for 13 abdominal organs; two lesion-wise metrics: DSC and NSD scores; two efficiency metrics: runtime and area under GPU memory-time curve. 
    \item Step 2. Rank algorithms for each of the 400 testing cases and each metric. Each algorithm has 2400 (400x6) rankings.
    \item Step 3. Compute the final rank for each algorithm by averaging all the rankings.
\end{itemize}

\subsection*{Ranking stability and statistical analysis}
% https://www.nature.com/articles/s41598-021-82017-6
We applied the bootstrapping approach and computed Kendall’s $\tau$~\cite{kendall-tal} to quantify the variability of the ranking scheme. 
Specifically, we first extracted 1000 bootstrap samples from the international validation set and computed the ranks again for each bootstrap sample. Then, the ranking agreement was quantified by Kendall’s $\tau$. 
Kendall’s $\tau$ computes the number of pairwise concordances and discordances between ranking lists. Its value ranges $[-1, 1]$ where -1 and 1 denote inverted and identical order, respectively. A stable ranking scheme should have a high Kendall’s $\tau$ value that is close to 1.
Wilcoxon signed rank test was used to compare the performance of different algorithms. Results were considered statistically significant if the $p-$value is less than 0.05.
The following packages were used in the analysis: ChallengeR~\cite{challengeR}, Python 3~\cite{python}, Numpy~\cite{harris2020numpy}, Pandas~\cite{reback2020pandas}, Scipy~\cite{virtanen2020scipy}, PyTorch~\cite{paszke2019pytorch}, and matplotlib~\cite{matplotlib}.

\subsection*{Data availability}
All the datasets have been publicly available on the challenge website \url{https://codalab.lisn.upsaclay.fr/competitions/12239}.

\subsection*{Code availability}
The code, method descriptions, and docker containers of the top ten teams are available at \url{https://codalab.lisn.upsaclay.fr/competitions/12239#learn_the_details-awards}.
% The evaluation code is available at \url{https://github.com/JunMa11/FLARE}.

\bibliographystyle{IEEEtran}
\bibliography{JunRef}

\end{document}